\documentclass[10pt,twocolumn,twoside]{IEEEtran}
\usepackage{amsmath,graphicx,cite}

\everymath{\displaystyle}

\usepackage{setspace}
\usepackage[mathcal]{euscript}
\usepackage{amssymb,amsmath}
\usepackage{verbatim}
\usepackage{graphicx}
\usepackage{longtable}
\usepackage{rotating}
\usepackage{multirow}
\usepackage{tabularx}
\usepackage{bigints}
\usepackage{tikz}
\usepackage{multirow} 
\usetikzlibrary{arrows,backgrounds,snakes}
\usepackage{amsmath,amstext,amsfonts,amssymb}
\usepackage{amsbsy}
\usepackage{amsthm}
\usepackage{diagbox}
\usepackage{mathabx}
\usepackage{color}
\usepackage{morefloats}
\usepackage{float}
\usetikzlibrary{arrows,shapes,positioning,shadows,trees}
\usetikzlibrary{decorations.pathreplacing}
\usepackage{colortbl}  
\definecolor{tabcolor}{rgb}{255,0,0} 
\usepackage{subcaption}

\tikzset{
  basic/.style  = {draw, text width=2cm, drop shadow, font=\sffamily, rectangle},
  root/.style   = {basic, rounded corners=2pt, thin, align=center,
                   fill=green!30},
  level 2/.style = {basic, rounded corners=6pt, thin,align=center, fill=white!60,
                   text width=10em},
  level 3/.style = {basic, thin, align=left, fill=pink!60, text width=6.5em}
}

\definecolor{darkred}{rgb}{0.5,0.0,0.0}
\definecolor{darkblack}{rgb}{0.2,0.1,0.6}

\newcommand{\ul}{\underline}

\newcommand{\tc}{\textcolor}

\newcommand{\Gm}{\textbf{G}}

\begin{document}

\title{Using Continuous Power Modulation for Exchanging Local Channel State Information}
\author{Chao Zhang, Samson Lasaulce, and Vineeth S. Varma}

\maketitle

\begin{singlespace}

\begin{abstract}
This letter provides a simple but efficient technique, which allows each transmitter of an interference network, to exchange local channel state information with the other transmitters. One salient feature of the proposed technique is that a transmitter only needs measurements of the signal power at its intended receiver to implement it, making direct inter-transmitter signaling channels unnecessary. The key idea to achieve this is to use a transient period during which the continuous power level of a transmitter is taken to be the linear combination of the channel gains to be exchanged. 
\end{abstract}

\begin{IEEEkeywords}
Distributed power control, interference channel, iterative water-filling algorithm, global channel state information, power domain channel estimation
\end{IEEEkeywords}

\section{Introduction}
\label{sec:intro}

To perform operations such as power control or interference coordination in distributed interference networks, a transmitter requires knowledge about the channel gains or qualities of the different links in presence. Aside from limitations such as \tc{black}{complexity}, one limitation in the implementation of globally efficient distributed power control policies, is the problem of channel state information (CSI) availability at the transmitter. Typically, global or network performance metrics depend on global CSI, i.e., all the channel gains in presence. While many estimation techniques or protocols to acquire local CSI (i.e., the channel gains of links from all transmitters to a given receiver) at a transmitter are available in literature \tc{black}{(see e.g., \cite{ozdemir-st-2007}\cite{boudreau-commmag-2009}),} there are not many works providing techniques to acquire global CSI. The existing techniques to acquire global CSI, typically rely on the existence of inter-transmitter signaling channels (see \cite{david}\cite{de2013csi}\cite{deghel2015queueing}), which may be unavailable in practice. Remarkably, it has been shown recently in \cite{varma-eusipco-2015} that global CSI\footnote{It is essential to note that the estimation techniques proposed in the present paper, and in \cite{varma-eusipco-2015}, differ from classical estimation techniques as they are performed in the power domain, and do not require training or pilot signals. While channel acquisition may take some time, note that regular communication is uninterrupted and occurs in parallel.} can be acquired from the sole knowledge of the received signal strength indicator (RSSI) or signal-to-interference plus noise ratio (SINR), and therefore making dedicated feedback and inter-transmitter channels unnecessary. The motivation for assuming such a feedback is given by power control schemes such as the iterative water-filling algorithms (IWFA) \cite{yu-jsac-2002}\cite{scutari-tsp-2009} which assume a transient or exploration period during which the power control algorithm updates the transmit power. For this, each transmitter updates its power by relying on individual SINR feedback. As shown in \cite{varma-eusipco-2015}, SINR feedback is sufficient to allow every transmitter to recover local CSI and exchange it by using the novel idea of power modulation, namely, the transmit power itself is used to embed information about the channel.   

In this letter, we present an alternative to the solution given in \cite{varma-eusipco-2015} to exchange local CSI. The latter solution involves quantizing local channel gains, power modulating the corresponding bits with discrete power levels, and using a lattice-type decoding scheme at each transmitter, which has to recover the power levels of the others. The CSI exchange technique in \cite{varma-eusipco-2015} has a least two advantages: it can be used in wireless systems where only discrete power levels are allowed (see e.g., \cite{sesia-book-2009}); it can be made robust to local CSI estimation noise and imperfections on the knowledge of the RSSI (or SINR). However, if continuous power levels are allowed, local CSI is well estimated, and the RSSI quality is good, a much simpler solution can be used namely, to use \textit{continuous} power modulation (CPM)\tc{black}{, which is the purpose of this letter. Technically, the technique proposed here distinguishes from \cite{varma-eusipco-2015} by the fact that it uses continuous power modulation (in a very specific way which allows local CSI exchange), and relies on time-sharing (which is especially relevant for power domain channel estimation). Additionally, the proposed technique has the advantage of having very low complexity.}

\begin{table}[h]
\begin{tabular}{|p{3cm}|p{5cm}|}
\hline
\multicolumn{2}{|c|}{{\textbf{Important notations}}}\\
\hline
\textbf{{Symbol}} & {\textbf{Meaning}}  \\
\hline
{$g_{ji}$} & {Channel gain from Transmitter $j$ to Receiver $i$ } \\
\hline
{$\widetilde{g}_{ji}$} & {Estimate of $g_{ji}$ available at Transmitter $i$}\\
\hline
{$\widehat{g}^k_{ji}$ }& {Estimate of $g_{ji}$ available at Transmitter $k \neq i$}  \\
\hline
{$\ul{g}_i = (g_{1i},...,g_{Ki})^{\mathrm{T}}$} & {Local CSI for Transmitter $i$}\\
\hline
{$\widehat{\ul{g}}_j^{i} = (g_{1j},...,g_{Kj})^{\mathrm{T}}$} &{Estimate of the local CSI of Transmitter $j$ available at Transmitter $i$}\\ 
\hline
{$\Gm$}& {Global channel matrix}\\
\hline
{$\widehat{\Gm}^k$} & {Estimate of  $\Gm$ available at Transmitter $k$}  \\
\hline
\end{tabular}
\vspace{-.01cm}
\end{table}

\section{Proposed technique}
\label{sec:proposed-technique}

We consider an interference channel with $K$ transmitter-receiver pairs which are assumed to operate on the same band \cite{david}; the multi-band scenario readily \tc{black}{follows} and is considered in the section dedicated to simulations (Sec.~\ref{sec:simulations}). The channel gain of the link between Transmitter $i \in \{1,...,K\}$ and Receiver $j \in \{1,...,K\}$ is denoted by $g_{ij} \geq 0$ and is assumed to lie in the interval $[g_{ij}^{\min}, g_{ij}^{\max}]$. \tc{black}{Channel gains are assumed to be constant over each transmitted data frame, a frame being composed of several time-slots. Each frame comprises two phases.} During the first $T\geq 1$ time-slots, transmit power levels $p_1$,...,$p_K$ vary from time-slot to time-slot and corresponding received signal (RS) power measurements are used to estimate \tc{black}{global CSI} (exploration phase), $p_i $ being subject to a classical power limitation $p_i \in [0, P_{\max}]$. Once the transmitters have acquired \tc{black}{a global CSI estimate}, they can find the fixed power level at which each of them will operate over all the remaining time-slots (exploitation phase). 

Let us denote by $\widetilde{g}_{1i},...,\widetilde{g}_{Ki}$ the local channel gain estimates available to Transmitter $i$. \tc{black}{As mentioned in Sec.~\ref{sec:intro}, local CSI can be acquired by using classical estimation techniques which, for example, assume that each transmitter sends to the receivers a pilot or training sequence in the signal domain.} One of the key features of the proposed technique is to adjust the power level of Transmitter $i$ \tc{black}{on time-slot $t \in \{1,...,T\}$},  as follows:
\begin{equation}
p_{i}(t)=\sum_{j=1}^{K} \overline{a}_{ji}(t) \widetilde{g}_{ji}
\label{eq:apm5}
\end{equation}
where
\begin{equation}
\overline{a}_{ji}(t)=a_{ji}(t)\frac{P_{\max}}{g_{ji}^{\max}}, \ a_{ji}(t) \geq 0, \ \sum_{j=1}^{K}a_{ji}(t)=1.
\end{equation}
Therefore, the power levels used during the exploration phase\footnote{Just as for the IWFA, tuning the power levels in the proposed way may induce some optimality loss, but in the scenarios of interest the influence of the exploration phase on the performance is negligible.} conveys information about local CSI. It turns out that local CSI information can be recovered, provided the interference is observed either through RSSI or SINR feedback. Here we consider RSSI feedback, which has the advantage of directly leading to linear estimators. In general, the RS power at Receiver $i$ on time-slot $t$ writes as follows:
\begin{equation}
\omega_i(t) = \sum_{j=1}^K g_{ji} p_j(t) + \sigma^2
\end{equation} 
where $\sigma^2$ is the receive noise variance. But in a real wireless system the RS power is quantized and transmitted through a noisy feedback channel (the corresponding quantizer and channel will be specified in Sec.~\ref{sec:simulations}). Thus Transmitter $i$ has only access to an observed or a noisy version of $\omega_i(t)$, which is denoted by $\widetilde{\omega}_i(t)$. To facilitate and make more accurate the local CSI exchange procedure, the used power levels during the exploration phase are imposed to follow a time-sharing rule. This means that during the exploration phase the power level of Transmitter $i$, is chosen either to follow (\ref{eq:apm5}) or to be zero. Assuming time-sharing for the exploration phase, {Transmitter $j$ is only active when} $t \in \{ t_j+1,t_j+2,\dots, t_j+K\}$ with $t_j:=j(K-1)$ . The observed RS power at Transmitter $i$ when Transmitter $j$ is active expresses as:
\begin{equation} \begin{array}{lll}
\widetilde{\omega}_{i}^j(t) & = & {\omega}_{i}^j(t)+z_{i,1}^j(t)\\
&=&{g}_{ji}  p_{j}(t) + \sigma^2+z_{i,1}^j(t)\\
&=& \widetilde{g}_{ji}  p_{j}(t) + \sigma^2+z_{i,1}^j(t)+z_{i,2}^j(t)\\
&=& \widetilde{g}_{ji}  p_{j}(t) + \sigma^2+z_{i}^j(t)
 \end{array}\label{eq:apm6}\end{equation}
where $z_{i,1}^j(t) =   \widetilde{\omega}_{i}^j(t)-
{\omega}_{i}^j(t)$ and $z_{i,2}^j(t)=\left(g_{ji} - \widetilde{g}_{ji} \right) p_{j}(t)$. By substituting $p_j(t)$ in (\ref{eq:apm6}) by its expanded version (\ref{eq:apm5}), it follows that the sequence of RS powers observed by Transmitter $i$ when Transmitter $j$ is active, expresses as:
\begin{equation}
\underline{\widetilde{\omega}}_{i}^j= \mathbf{P}^j \underline{\widetilde{g}}_{j} \widetilde{g}_{ji} + \underline{z}_{i}^j + \sigma^2 \underline{1}
\end{equation}
where $\underline{\omega}_{i}^j = (\omega_i^j(t_j+1),..., \omega_i^j(t_j+K) )^{\mathrm{T}} $, $\underline{\widetilde{g}}_{j}  =(\widetilde{g}_{1j},...,\widetilde{g}_{Kj})^\mathrm{T} $, $\underline{z}_{i}^j =( z_{i}^j(t_j+1),...,z_{i}^j(t_j+K))^{\mathrm{T}}$, $ \underline{1} = (1,1,...,1)^{\mathrm{T}} $, and 
\begin{equation}
\mathbf{P}^j= P_{\max}
\left(
\begin{array}{ccc}
\frac{a_{1j}(t_j+1)}{g_{1j}^{\max}}& \dots &\frac{a_{Kj}(t_j+1)}{g_{Kj}^{\max}} \\
  \vdots  & \vdots & \vdots \\
 \frac{a_{1j}(t_j+K)  }{g_{1j}^{\max}}& \dots &\frac{a_{Kj}(t_j+K)}{g_{Kj}^{\max}} 
 \end{array}
\right).
\label{eq:apm7}
\end{equation}
This means that when $j$ is broadcasting from time $t_j+1$ to $t_j+K$, the sequence of transmit powers it uses (as a column vector), over the $K$ time slots, is given by $\mathbf{P}^j\underline{\widetilde{g}}_{j}$. Finally, the local CSI estimate $\widetilde{g}_{j}$ is estimated at Transmitter $i$ as:
\begin{equation}
\widehat{\underline{g}}_{j}^{i} = \frac{{\mathbf{P}^j}^{-1}}{\widetilde{g}_{ji} } \left( \underline{\widetilde{\omega}}_{i}^j -   \sigma^2 \underline{1}  \right).
\label{eq:ls}
\end{equation}\tc{black}{where $\widehat{g}^i_{kj}$ denotes the estimate of the channel $g_{kj}$ by transmitter $i$.}
Of course, writing the above implicitly assumes that the coefficients $a_{ji}$ are chosen properly. Note that more advanced estimators such as the minimum mean square error (MMSE) or maximum likelihood (ML) estimators might be used but here, low complexity is prioritized. Normal requirements in terms of local CSI and RSSI qualities are the targeted operating conditions for the proposed technique. \tc{black}{Once a global CSI estimate is available, it becomes possible for Transmitter $i$, $i \in \{1,...,K\}$, to maximize a common network performance criterion under the form}
\begin{equation}
\tc{black}{u(p_1,...,p_K; \widehat{\Gm}^i),}
\end{equation}
\tc{black}{ \tc{black}{$\widehat{\Gm}^i=\left[\widehat{\underline{g}}_{1}^{i}\cdots\widehat{\underline{g}}_{K}^{i}\right]$} being the global channel matrix estimate (this setup has been coined for distributed CSI and studied in \cite{dekerret-tit-2012}).}

\tc{black}{\section{Comparison of the proposed technique with \cite{varma-eusipco-2015}}}
\label{sec:simulations}

\tc{black}{\subsection{Technical comparison}}

\tc{black}{Compared to the local CSI exchange technique provided in \cite{varma-eusipco-2015}, the proposed technique has two distinguishing technical features. First, the transmit power during the exploration phase is continuous. Thus, the proposed technique can be seen as a complementary technique to \cite{varma-eusipco-2015} for scenarios in which discrete power levels are not allowed. Additionally, the continuous power is chosen in a very specific manner, namely to be the linear combination of the channel gains. Second, only one transmitter is active at a time during the local CSI exchange phase, which makes the estimation procedure simple, but it is observed to be very efficient via simulations. To understand the underlying problem let us consider a special case $K=3$ and Transmitter $1$. When three users are active at a time and $g_{31} >> g_{21}$ it becomes difficult to recover $p_2$ from $\omega_1(t) = g_{11}p_1(t) + g_{21}p_2(t)+g_{31}p_3(t)+\sigma^2$. One drawback for only activating one transmitter at a time appears if only SINR feedback is available instead of RSSI feedback. In the presence of SINR feedback, at least two users have to be active at a time to allow information exchange in the power domain.}


\tc{black}{\subsection{Comparison of CSI exchange phase}}

If one assumes that the number of required observations has to be equal to the number of unknowns to estimate, the local CSI exchange technique of \cite{varma-eusipco-2015} requires $K(K-1)$ time-slots. During this CSI exchange phase, regular communication occurs in parallel, but with potentially high interference. Under the same assumption, the technique proposed here requires $T=K^2$ time-slots for the exchange phase since each transmitter has $K$ channel gains to be exchanged and only one user is active at a time (regular communication is effectively time-division multiple access). For $K$ being respectively equal to $2$, $3$, and $4$, this corresponds to an additional cost in terms of time-slots of $100\%$, $50\%$, and $33\%$. {Indeed, the number of effectively interfering users using the same channel (meaning operating on the same frequency band, at the same period of time, in the same geographical area) will typically be small in practice and not exceed $3$ or $4$, which makes the number of time-slots of the exploration phase reasonable.}

\tc{black}{\subsection{Simulations}}

As mentioned in Sec.~\ref{sec:proposed-technique}, we consider the multi-band case. Here, $s$ will stand for the band index, and $S$ for the number of bands. We assume, for ease of exposition, that the channel gain statistics (path losses) are symmetric over all the bands; which is why in some places, the band index is removed. To be able to compare the technique provided in this letter to the state-of-the-art technique given in \cite{varma-eusipco-2015}, we consider two performance metrics, namely, the estimation signal-to-noise ratio (ESNR) and the sum-rate, which are defined as follows: 
\begin{equation}\label{eq:ESNR}
\mathrm{ESNR}_i= \frac{\mathbb{E}[\|\mathbf{G}\|^2]}{\mathbb{E}[\|\mathbf{G}-\widehat{\mathbf{G}}_i\|^2]} 
\end{equation}
where $\|.\|^2$ stands for the Frobenius norm and $\mathbf{G} = [\mathbf{G}^1 \cdots \mathbf{G}^S]$ is the global channel matrix and the entries of $\mathbf{G}^s$ are the channel gains for "band $s$" $g_{ij}^s$, $i$ and $j$ being respectively the row and column indices. The matrix $\widehat{\mathbf{G}}_i$ corresponds to the estimate which is available to Transmitter $i$. The sum-rate is defined as: 
\begin{equation}\label{eq:sum-EE-def}
u(\ul{p}_1,...,\ul{p}_K;\Gm) =  \sum_{s=1}^S \sum_{i=1}^K 
\log_2(1+\mathrm{SINR}_i^s(p_1^s,...,p_K^s; \mathbf{G}^s  )) 
\end{equation}
where: $p_i^s$ is the transmit power level for band $s$, $\mathrm{SINR}_{i}^s(t)   = \frac{ g_{ii}^s  p_{i}^s(t)}{\displaystyle{ \sigma^2+\sum_{j \neq i} g_{ji}^s  p_{j}^s(t) }}$.
We assume that RS power measurements are quantized uniformly in a dB scale with $N$ bits and the quantizer input dynamics or range in dB is $[\mathrm{SNR(dB)}-20, \mathrm{SNR(dB)}+ 10]$ where $
\mathrm{SNR(dB)}=10 \log_{10}\left(  \frac{P_{\max}}{\sigma^2} \right) 
$. The quantizer used by Receiver $i$ produces labels which are sent to Transmitter $i$ through a feedback channel and correctly received with probability $1-\varepsilon$ with $0 \leq \varepsilon \leq 1$. We also need to define the signal-to-interference ratio (SIR) for the case of $K=2$ transmitters: 
\begin{equation}
\mathrm{SIR(dB)} = 10 \log_{10}\left(  \frac{\mathbb{E}(g_{11})}{\mathbb{E}(g_{21})} \right) = 10 \log_{10}\left(  \frac{\mathbb{E}(g_{22})}{\mathbb{E}(g_{12})} \right) 
\end{equation}
which can be written when the channel statistics are symmetric (which we assume to be true in our simulation settings). We chose the parameters $g_{ij}^{\min} = 0.01 \mathbb{E}(g_{ij})$, $g_{ij}^{\max} = 5 \mathbb{E}(g_{ij})$ (when $i \neq j$) and $\mathbb{E}(g_{ii})=1$ for all the simulations in our setting, with $\mathbb{E}(g_{ij})$ determined by the SIR indicated. The channel gain dynamics $\frac{g_{ij}^{\max}}{g_{ij}^{\min}}$ is thus equal to $27$ dB, which is a quite typical value in real systems. The ESNR is obtained by averaging over $10^4$ realizations for the channel matrix, the channel gains being chosen independently and according to an exponential law $\phi_{ij}( g_{ij} )  = \frac{1}{\mathbb{E}(g_{ij})}  \exp\left(- \frac{g_{ij}}{\mathbb{E}(g_{ij})}\right)$ (that is, Rayleigh fading is assumed). The local CSI estimates are assumed to be perfectly known.

\textbf{Influence of the choice of the parameters $a_{ij}(t)$.} First, we study the impact of the choice of the matrix $\mathbf{P}^j$ on the ESNR. Denote by $\mathbf{A}^j$ the matrix whose entries are the coefficients $a_{ij}(t)$ for all $t \in \{t_j+1,\dots,t_j+K\}$, i.e.,
\begin{equation}
\mathbf{A}^j= \left(
\begin{array}{ccc}
a_{1j}(t_j+1) & \dots & a_{Kj}(t_j+1)\\
  \vdots  & \vdots & \vdots \\
 a_{1j}(t_j+K) & \dots & a_{Kj}(t_j+K) 
 \end{array}
\right).
\end{equation}

{Interestingly, for typical scenarios, the particular choice $\mathbf{A}^j=\mathbf{I}$ only induces a quite small performance loss with respect to the optimal choice e.g., measured in terms of ESNR. In this respect, we perform simulations to compare the ENSR obtained by choosing the best possible $\mathbf{A}^j$ over that of choosing $\mathbf{A}^j=\mathbf{I}$, for all $j$. For $S=1$, $K=2$, $N=8$, $\varepsilon=1\%$, $\mathrm{SNR(dB)}=30$, and $\mathrm{SIR(dB)}=10$ our comparison has shown that choosing the best matrix only provides marginal improvements. Indeed, for typical values for the SIR (say above $5$ dB), estimating the cross-channel gains reliably lead to matrices which are quite similar to the identity matrix; otherwise, the influence of the cross-channel gains in the sum (1) or in the RS power might be dominated by that of the direct channel. Therefore, for the rest of this letter, we will choose $\mathbf{A}^j= \mathbf{I}$ for all $j$, as this choice results in a very low complexity technique and guarantees the invertibility of the power matrix in (7).}

\begin{figure}[h]
   \begin{center}
        \includegraphics[width=95mm,height=60mm]{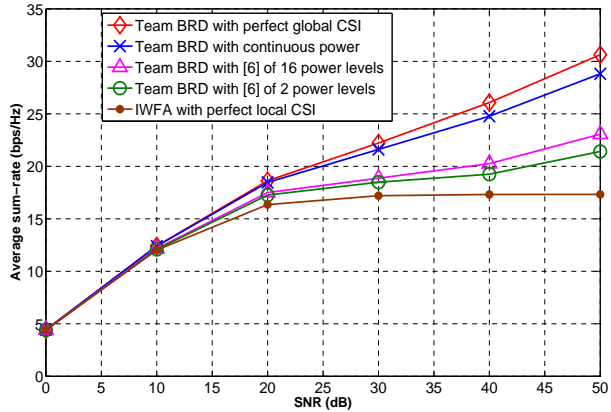}
    \end{center}
 \caption{\small Comparison of average sum-rate obtained by using the proposed technique (continuous power modulation) with \cite{varma-eusipco-2015}. We observe that our technique results in an average sum-rate that is very close to the ideal case of perfect global CSI.}
   \label{fig:sumrate}
\end{figure}

\textbf{Proposed scheme versus state-of-the-art.} We compare the performance in terms of sum-rate of our proposed technique, with the state-of-the-art techniques in Fig. \ref{fig:sumrate}. We consider the parameters $K=4$ (number of users), $S=2$ bands, perfect \emph{local} CSI estimate and $\mathrm{SIR(dB)=10}$. The RSSI feedback is assumed to be with $N=8$ bits and $\varepsilon=0.01$. We compare the performance measured by the sum-rate using CSI exchange using our proposed scheme, of {CPM}, with that \tc{black}{of: 1) perfect global CSI (ideal case); 2)local CSI exchange using a power modulation based on Lloyd-Max quantization (as in \cite{varma-eusipco-2015}) with $2$ or $16$ quantization levels; 3) the iterative water-filling algorithm (IWFA).} When global CSI is available (perfect or otherwise), we implement a \textit{team best response dynamics (BRD)}\footnote{Team BRD for sum-rate with power control implies that each transmitter iteratively updates its best transmit power given the other transmit powers until all the powers converge. Since global CSI is available, each transmitter can do this offline by assuming an arbitrary initial power vector and with perfect global CSI, all transmitters will converge to the same equilibrium.} to select the power control, where each transmitter uses the CSI available for the BRD. In the case of IWFA, no exchange of local CSI is required (no phase 2), but there is a time taken for the algorithm to converge. This figure demonstrates the performance improvement offered by our proposed modulation technique in terms of sum-rate, and we can observe that the team-BRD with {CPM} achieves a sum-rate that is very close to that with perfect global CSI (which is the ideal case). 

\begin{figure}[h]
   \begin{center}
        \includegraphics[width=95mm,height=60mm]{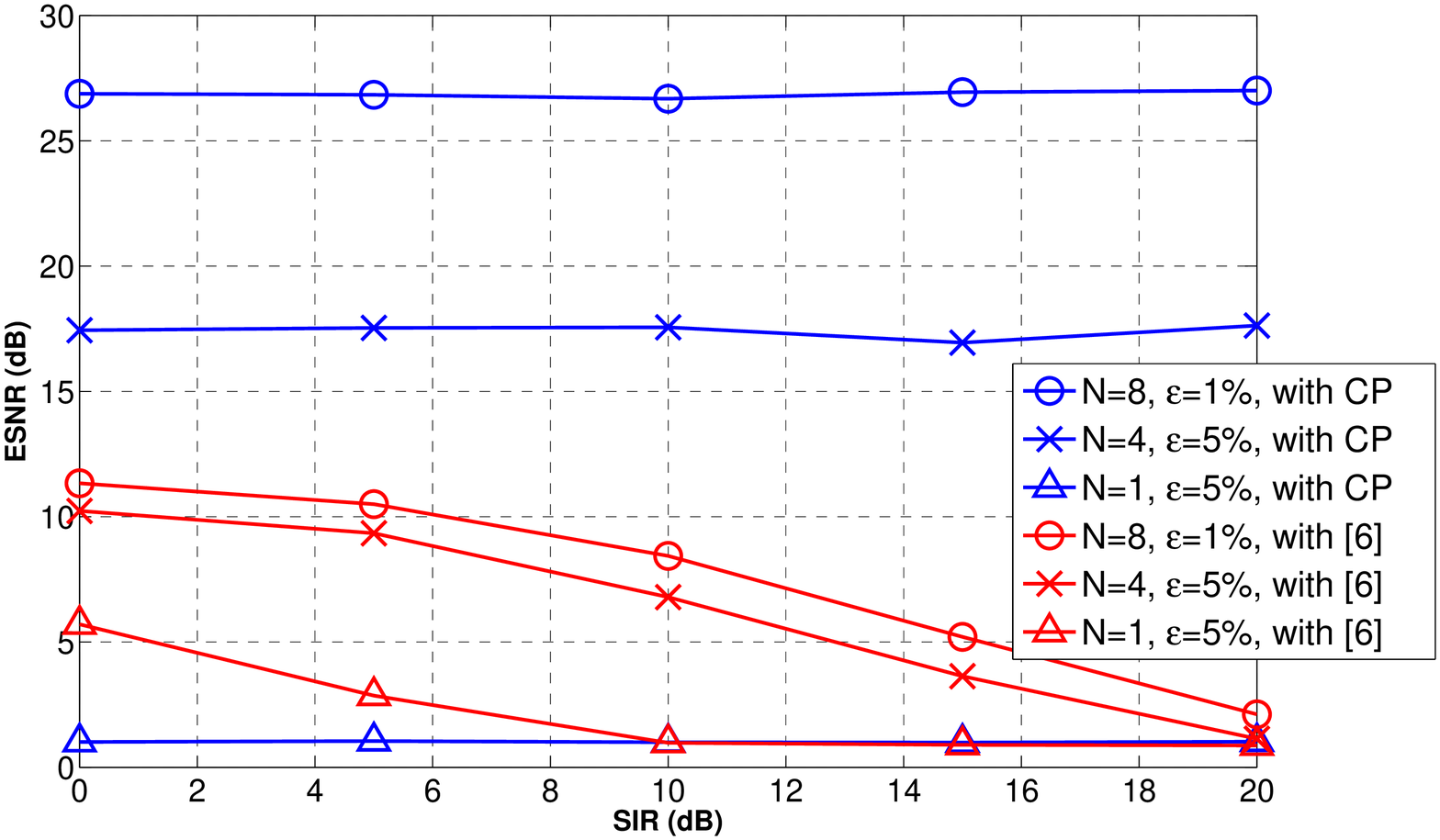}
    \end{center}
 \caption{\small ESNR against SIR. Using continuous power modulation to exchange local CSI appears to be a relevant choice when the RSSI quality is good or even medium. Under severe conditions (e.g., when only an ACK/NACK-type feedback is available for estimating the channel), quantizing the channel gains and power modulating the corresponding labels as in \cite{varma-eusipco-2015} is more appropriate.}
 	    \label{fig:ESNR_noise}
\end{figure}

Fig.~\ref{fig:ESNR_noise} represents the ENSR in dB against the SIR in dB for
which represents the sum-EE as a function of the SNR and assumes a similar setting to Fig. \ref{fig:sumrate} except that here, $\mathrm{SNR(dB)}=30$, $K=2$ users and $S=1$ bands. Three scenarios are considered: $(N,\varepsilon) = (8, 1 \%)$, $(N,\varepsilon) = (4, 5 \%)$, and $(N,\varepsilon) = (1, 5 \%)$. The first scenario corresponds to typical conditions in terms of quality for the RSSI, while the two others correspond to quite severe conditions. The scenario with only one RS power quantization bit can be seen as a scenario with an ACK/NACK feedback. For each scenario, two schemes are compared: the scheme proposed in this letter and the one used in \cite{varma-eusipco-2015} which relies on channel gain quantization (here with $1$ or $4$ bits), discrete power modulation with $L$ levels and lattice decoding. {In Fig.~\ref{fig:ESNR_noise} we compare with the case of $L=2$.}

Fig.~\ref{fig:ESNR_noise} clearly shows that the proposed technique provides a performance in terms of ESNR, which is independent of the SIR level; this is one of the effects of using time-sharing in the exploration phase. When the quality of the RSSI is good, it is seen that the proposed technique provides a very significant gain in terms of ESNR; the gain ranges from $10$ dB to $25$ dB, depending on the SIR level. It is only when the RSSI quality is severely degraded (namely, when $(N,\varepsilon) = (1, 5 \%)$) that the proposed technique does not perform well when compared to the technique of \cite{varma-eusipco-2015} when using $L=2$ (as seen in Fig.~\ref{fig:ESNR_noise}).

\end{singlespace}

\section{Conclusion}
\label{sec:conclusion}

The two main features of the proposed technique are using continuous power modulation in a very specific manner, and to activate one transmitter at a time. Although continuous power modulation may be thought to be sensitive to imperfect RSSI feedback and imperfect local CSI, it is shown to perform very well under normal conditions in terms of RSSI quality.

\color{black}

\end{document}